\documentclass[10pt,journal,letterpaper]{IEEEtran}
\IEEEoverridecommandlockouts
\usepackage[T1]{fontenc}
\usepackage{orcidlink}
\usepackage{setspace}

\IEEEoverridecommandlockouts
\usepackage{fancyhdr}
\usepackage[numbers, sort&compress]{natbib}

\usepackage{float}
\usepackage{enumerate}
\usepackage{hyperref}
\usepackage{amsmath,amssymb,amsfonts}
\usepackage{algorithmic}
\usepackage[ruled,vlined]{algorithm2e}
\usepackage{graphicx}
\usepackage{array}     
\usepackage{amsmath}   
\usepackage{textcomp}
\usepackage{xcolor}
\usepackage{physics}
\usepackage[font=footnotesize, skip=0pt]{caption}
\usepackage{subcaption}
\usepackage[percent]{overpic} 

\usepackage{comment}

\usepackage{makecell}

\setcellgapes{2pt}
\makegapedcells

\usepackage{booktabs}
\usepackage{multirow}
\def\BibTeX{{\rm B\kern-.05em{\sc i\kern-.025em b}\kern-.08em
    T\kern-.1667em\lower.7ex\hbox{E}\kern-.125emX}}
\begin{document}
\bstctlcite{MyBSTcontrol}
\title{Combined Quantum and Post-Quantum Security Performance Under Finite Keys\\}
\author{Aman Gupta$^{*}$, Ravi Singh Adhikari$^{*}$, Anju~Rani$^{*}$, Xiaoyu~Ai$^{*}$,  and~Robert~Malaney$^*$ 
\thanks{$^*$Aman Gupta, Ravi Singh Adhikari, Anju Rani, Xiaoyu Ai, and Robert Malaney are with the School of Electrical Engineering and Telecommunications, University of New South Wales, Sydney, Australia.}
}

\maketitle

\begin{abstract}
Recent advances in quantum-secure communication have highlighted the value of hybrid schemes that combine Quantum Key Distribution (QKD) with Post-Quantum Cryptography (PQC). Yet most existing hybrid designs omit realistic finite-key effects on QKD key rates and do not specify how to maintain security when both QKD and PQC primitives leak information through side-channels. These gaps limit the applicability of hybrid systems in practical, deployed networks. In this work, we advance a recently proposed hybrid QKD-PQC system by integrating tight finite-key security to the QKD primitive and improving the design for better scalability. This hybrid system employs an information-theoretically secure instruction sequence that determines the configurations of different primitives and thus ensures message confidentiality even when both the QKD and the PQC primitives are compromised. The novelty in our work lies in the implementation of the tightest finite-key security to date for the BBM92 protocol and the design improvements in the primitives of the hybrid system that ensure the processing time scales linearly with the size of secret instructions.

\end{abstract}

\begin{IEEEkeywords}
Quantum key distribution, finite-key security, quantum cryptography, quantum resistant, post-quantum cryptography, hybrid cryptography.
\end{IEEEkeywords}

\section{Introduction}
The rapid adoption of cryptographic designs that are resistant to quantum attacks is a pressing requirement in current communication systems. Although fully scalable quantum computers capable of breaking existing cryptographic schemes~\cite{proctor2025benchmarking} have not yet been realized, it remains crucial to transition to quantum-resistant\footnote{The term `quantum-resistant' describes cryptographic systems that are secure against known quantum algorithmic attacks, such as Shor's and Grover's algorithms, ensuring that system security is not compromised by quantum computers within polynomial time complexity.} schemes due to the concept of harvest-now-decrypt-later~\cite{10478407}. There are two quantum-resistant solution fields, post-quantum cryptography (PQC); and quantum cryptography, mainly quantum key distribution (QKD). PQC schemes are based on mathematically hard problems that even known quantum algorithms cannot compromise~\cite{liu2024post}. They are easy to integrate with existing infrastructure and have been extensively studied. However, PQC is still relatively new and may reveal unforeseen vulnerabilities, particularly against side-channel attacks. Side-channel attacks on PQC schemes still remain an open problem~\cite{hoffmann2023polka}. On the other hand, QKD protocols are developed to a great extent, such that many QKD implementations are now commercially available and have been deployed in some real-time communication applications~\cite{xu2020secure, makarov2024preparing, kish2025quantumkeydistribution}. QKD, in principle, offers information-theoretic security; however, it is challenging to integrate with existing cryptographic infrastructure, and a trade-off between the QKD key rate and its security often arises when choosing from different QKD protocols. There are some implementations of QKD protocols~\cite{10225641, zhang2022experimental,zhou2025experimentalsidechannelsecurequantumkey,jiang2025practicalissuessidechannelsecurequantum} that are considered robust against most known side-channel attacks but suffer from very low key rates. Some others achieve higher key rates but are vulnerable to known side-channel attacks~\cite{biswas2021experimental,pantoja2024electromagnetic,arteaga2022practical}. 
Consequently, hybrid cryptographic architectures integrating QKD, PQC, and classical schemes\footnote{We use the phrase `classical schemes' to mean the key sharing algorithms that are deployed in the current infrastructure, prior to PQC algorithms. The algorithms, such as RSA and ECC, fall in this category.} represent the future of secure communications, offering adaptability to evolving threats and leveraging the unique advantages of each primitive to achieve long-term resilience, for example~\cite{aquina2024quantum, zeng2024practical,rani2025obfuscatedquantumpostquantumcryptography}. Most hybrid cryptographic designs combine the keys generated through QKD and PQC primitives with classical keys to generate hybrid key material~\cite{garms2024experimental, bindel2019hybrid, ricci2024hybrid, dowling2020many, bruckner2023end}, while others propose the use of PQC to provide quantum-resistant security to classical communications involved in QKD~\cite{ghashghaei2024enhancing, djordjevic2020joint, wang2021experimental}. Furthermore, the work of~\cite{9390172} proposes the use of QKD keys as a seed in PQC algorithms. 

While hybrid QKD-PQC systems are among the most promising solutions towards achieving quantum-resistant cryptographic architectures, their security still depends on the assumption that at least one of the constituent primitives remains uncompromised. If independent side-channel vulnerabilities were to be exploited simultaneously in both QKD and PQC components, such combined systems would fail to provide end-to-end confidentiality. Additionally, previous hybrid QKD–PQC systems overlook the implementation vulnerabilities in the encryption schemes itself. 

To address the aforementioned vulnerabilities, in our previous work~\cite{rani2025obfuscatedquantumpostquantumcryptography}, we proposed and experimentally demonstrated a hybrid QKD-PQC system, hereafter termed the hybrid obfuscated quantum system (HOQS). This hybrid system established symmetric QKD keys via the BBM92 protocol, assuming an asymptotic security framework, asymmetric PQC keys using the Crystals-Kyber, a post-quantum public key encryption (PQ-PKE) scheme. The $\text{HOQS}$ also shared information-theoretically (IT) secure instruction sequences~(ISs) between two parties (say Alice and Bob). 
The IS was encrypted by XORing it with a subset of a pre-shared key (PSK) prior to transmission, making it resistant to side-channel attacks. 

Here, we extend our previous work~\cite{rani2025obfuscatedquantumpostquantumcryptography}, with the following  contributions:
 (i) We demonstrate for the first time how the tightest bounds on the finite QKD key rates can be deployed in a working QKD system. (ii)
     We introduce a range of modifications in $\text{HOQS}$ in terms of key sharing and encryption primitives to prevent rapid ciphertext growth and subsequent poor scaling performance, and also to eliminate potential vulnerabilities to re-encryption attacks~\cite {cryptography7040049}. 
    (iii) We experimentally demonstrate the feasibility and scaling of our modified system in terms of processing time compared to some implementations of $\text{HOQS}$.
Henceforth, we refer to our modified system as $\text{HOQS+}$  (or simply as the `hybrid system').

The remainder of this paper is organized as follows.  In Section~\ref{Sec: system_model}, we present the details of the modifications we implemented in $\text{HOQS}$ to develop the resulting $\text{HOQS+}$. In Section~\ref {Sec: finiteQKD} we describe the implementation of the finite-key framework of the QKD primitive. Section~\ref{Sec: results} presents our results on the QKD key rates and the performance of the delivered scalability.
In Section~\ref{Sec: conclusion}, we provide the conclusions of our work. 

\section{System model\label{Sec: system_model}}
In this section, we discuss the modifications introduced in the PQC and the hybrid encryption (HE) primitives of the HOQS in order to improve its scalability\footnote{We note that in our previous work, we proposed a hybrid QKD-PQC system. Many set-ups in that system can lead to acceptable scaling performance, but some important set-ups can lead to poor scaling performance, such as the specific set-up used for illustration in~\cite{rani2025obfuscatedquantumpostquantumcryptography}.  In this work, we present a different implementation that delivers vastly improved scaling performance for that specific set-up. All comparisons to HQOS in this work refer to that specific poor-scaling setup.}. Before discussing the modifications that we applied to $\text{HOQS}$ to develop $\text{HOQS+}$, we briefly describe the operational design of the former.  
\subsection{HOQS\label{sec:System A}}
\begin{figure}
    \centering
    \includegraphics[width=\linewidth]{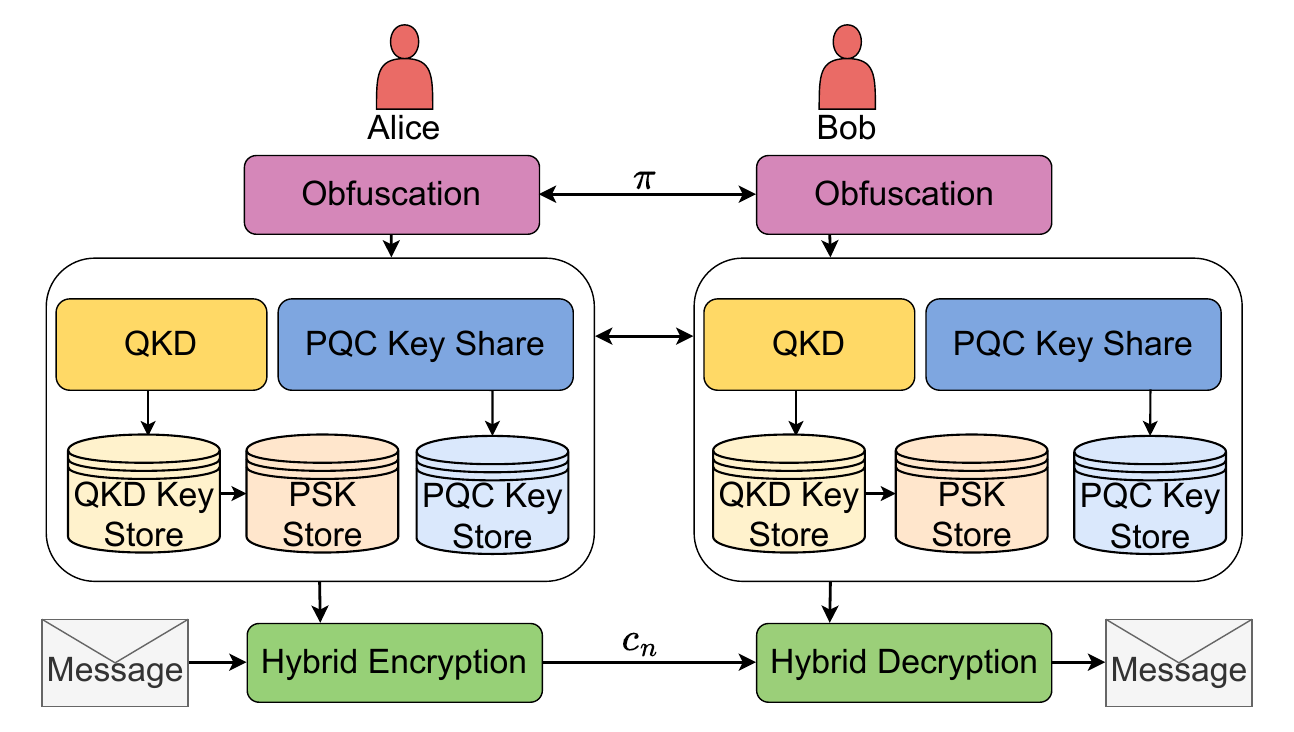}
    \caption{A general operational overview of the proposed hybrid QKD-PQC system (common for both the $\text{HOQS}$ and the $\text{HOQS+}$). The encrypted IS is represented as $\pi$. The QKD and PQC primitives establish the corresponding QKD and PQC keys in real-time. In the $\text{HOQS}$, the PQC Key Share is a component used to establish an asymmetric key, and HE involves PKE, OTP, and AES encryption schemes. In the $\text{HOQS+}$, the PQC Key Share establishes symmetric PQC keys via KEM, and HE involves Ascon encryption with PQC keys, modified AES with $256$ bits of the PSK, and OTP with the QKD keys.}
    \label{fig:system_flow}
\end{figure}

A single cycle of both the HOQS and the HOQS+ is described as follows. (i)~An IS that obfuscates the operational configurations of the system is encrypted by XORing it with a subset of the PSK\footnote{A pre-shared key (PSK) is a finite set of IT secure bits that are shared between Alice and Bob \textit{a priori}. A PSK is an assumed resource in all IT-secured QKD protocols, and we employ subsets of the same PSK (deployed for QKD purposes) in other primitives of our hybrid system.} and then sent to Bob. Based on the IS, the configurations for the QKD post-processing, the PQC key sharing, and the HE primitives are determined. (ii)~The QKD process involves real-time synchronization of the timestamps of the received entangled photons. (iii)~Then, sifting, error-correction, and privacy amplification (QKD post-processing) are performed through the authenticated classical channel, where authentication is performed using the Wegman-Carter message authentication code (MAC) with a subset of the PSK. Another $256$~bit of the PSK is used as a key for the encryption of classical data by the advanced encryption standard (AES). Till this step marks one QKD session. (iv)~Subsequently, the PQC primitive uses Crystals-Kyber as a PKE scheme that generates a public-private key pair for Alice and Bob. (v)~Finally, the message to be communicated is encrypted with an HE scheme. This step marks the completion of one cycle of the hybrid system. The HE involves multiple cascades of encryptions with different encryption schemes, with their corresponding keys. Fig.~\ref{fig:system_flow} represents the operational components and the interactions between different primitives of the hybrid system, which is common to both HOQS and HOQS+. 

The IS also dictates the configuration of HE. For example, one instance of an IS could inform Alice and Bob about the number of times different encryption schemes are used and their order to be used by the HE function. Mathematically, IS $= (\xi_1,\xi_2,\cdots,\xi_i,\cdots)$, where $\xi_i \in \{\text{OTP, AES, PQ}\}$, represents an encryption scheme. The OTP refers to XOR encryption of an input with QKD keys, PQ refers to Crystals-Kyber PKE, and AES refers to the AES encryption in the counter mode (denoted simply by AES in this work). 

The enhancement in the security of the HOQS against simultaneous side-channel attacks arises from its independence between each key generation and encryption primitives and from its IT secure instruction sequence-based HE. Unlike conventional cascaded or multiple encryption schemes that are susceptible to meet-in-the-middle~\cite{gavzi2009cascade} or re-encryption~\cite{cryptography7040049} attacks, the HE in HOQS prevents both attacks via an IT secure IS and by ensuring that no two consecutive encryption primitives are the same. With $2^{\hat{N}_{obs}}$ distinct possible permutations of the unique ISs, even if all established keys from QKD and PQC are individually exposed through side-channel leakage, Eve's effective key space search is still upper bounded by $\mathcal{O}(2^{(\hat{N}_{obs}/4)(N_{\mathrm{AES}}+2)})$ operations, to compromise the confidentially of a message, where $N_{\mathrm{AES}}$ denotes the number of keys used for AES encryption and the $\hat{N}_{obs}$ is a parameter that represents the security strength. Hence, the system, HOQS and by extension HOQS+, preserves message confidentiality under simultaneous compromise of multiple primitives. The security of both the systems is directly proportional to a parameter $\hat{N}_{obs} \in \{0, 2,3,4,\cdots\}$. A more detailed discussion of these security arguments is provided in ~\cite{rani2025obfuscatedquantumpostquantumcryptography}.

\begin{figure}[t]
         \centering
         \includegraphics[width=\linewidth]{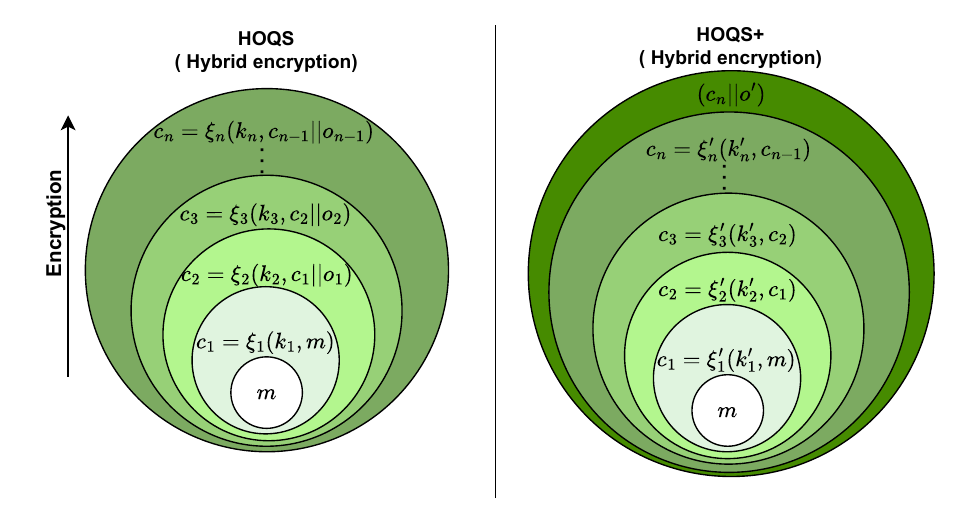}
         \caption{Architectures of the HE (cascaded) primitive within the hybrid QKD-PQC systems. On the left, the HE primitive of the $\text{HOQS}$, in which the output intermediate ciphertext, $c_i$ is concatenated with other relevant data, $o_i$, such as nonces, paddings, and tags. Here, $\xi_i \in$ \{OTP, AES, PQ\} represents the different encryption schemes with their corresponding keys, $k_i$. On the right, the HE primitive of the $\text{HOQS+}$ is shown, where the relevant data $o'$ is concatenated to the final ciphertext. In the $\text{HOQS+}$, $\xi'_i \in$ \{OTP, AES, Ascon\} with their corresponding keys, $k_i'$. The details on the derivation of $o_i$ for intermediate encryption and the encryption schemes $\xi'_i$ are given in Sec.~\ref{AES} and ~\ref{sec: Ascon}.}
         \label{fig:Cascade_encryption}
     \end{figure}

Each encryption scheme in the $\text{HOQS}$ is implemented based on the recommended standard practices, e.g., for each intermediate encryption, if the scheme is AES, treat the input as a single block, generate a random nonce and concatenate it to the end of the ciphertext~\cite{10.1007/978-3-031-15985-5_10}. Fig.~\ref{fig:Cascade_encryption} (left) shows the HE architecture of the $\text{HOQS}$. 
However, the usage of standard practices of individual encryption schemes in a HE primitive impacts the scaling performance of the $\text{HOQS}$ at large $\hat{N}_{obs}$. This is because the concatenated nonce also becomes a part of the input ciphertext to the next encryption. Additionally, each PKE operation generates an output approximately $m-$fold ($m \approx \mathcal{O}(\exp \small(\hat{N}_{obs}\small))$) larger than the input. At large $\hat{N}_{obs}$, this amplification in intermediate ciphertext size results in much longer execution times of the $\text{HOQS}$ compared to the $\text{HOQS+}$ (see Sec.~\ref{Sec: results}) than the expected theoretical model of $\lceil\hat{N}_{obs}/2\rceil$ proposed in our previous work.

\subsection{HOQS+\label{sec:HOQS+}}
\begin{figure*}[!hbt]
         \centering
         \includegraphics[width=0.8\linewidth]{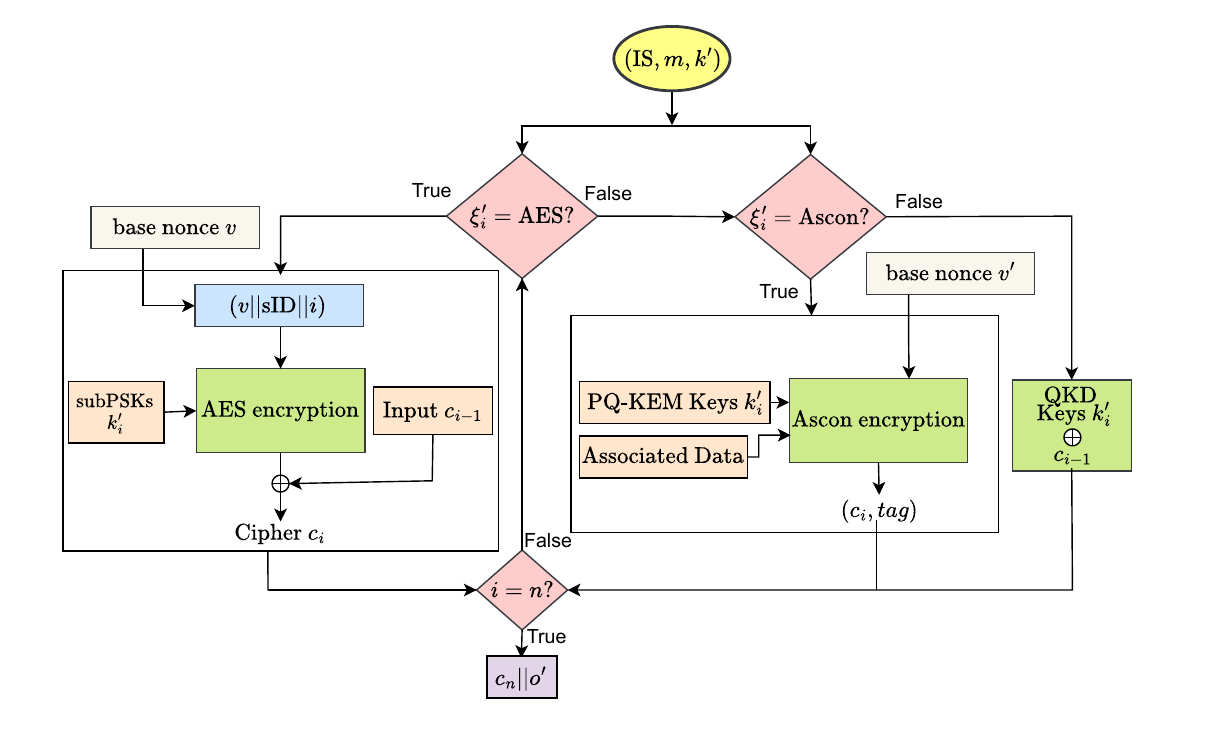}
         \caption{The data flow in the HE primitive of the proposed $\text{HOQS+}$. The inputs $\text{IS},m,k'$ represent the instruction sequence, the input message, and the complete set of keys (including QKD keys, PQC keys, and PSK subsets), respectively. The subPSKs denote a subset of pre-shared keys. It is assumed that the base nonces $v$ and $v'$ are generated using true random seeds. The counter block for AES is generated by concatenating the base nonce, $v$, with the sID and counter, $i$. Similarly, in Ascon encryption, the base nonce $v'$ comes from the part of the PQC keys concatenated with the counter. The string $o'$ is constructed by concatenating sID, base nonce $v$, associated data, and the padding size. Associated data is a public data used in Ascon encryption.}
         \label{fig:HE}
\end{figure*}

Next, we present the modifications to the $\text{HOQS}$, more specifically, its HE and PQC primitives. We use the individual encryption schemes more efficiently in the context of multiple cascaded encryption on a single message $m$ and use a key encapsulation mechanism (KEM) to establish the symmetric PQC keys instead of using PKE to establish asymmetric PQC keys. The flow of input message in the HE scheme in the $\text{HOQS+}$ is shown in Fig.~\ref{fig:HE}. The primary modifications introduced in the $\text{HOQS+}$ are summarised as follows.

\subsubsection{AES encryption in the HE\label{AES}}
In AES encryption-counter mode, which forms one of the encryption algorithms of the HE, it is required that the counter block-key pair be unique to satisfy the security arguments of AES. A counter block is constructed by concatenating a random base nonce, $v$, a session ID (sID), and a counter ($i$). The base nonce is generated only once per cycle of the system, and remains common to all the AES encryptions within that cycle. A unique sID is associated with each cycle number and the counter associated with the order of encryption; this ensures the uniqueness of the counter block. 
The random nonce is now only appended once at the end of the final ciphertext. 

\subsubsection{Ascon encryption in the HE\label{sec: Ascon}}
As noted above, the PKE component can introduce significant computational overhead in an HE due to the rapid increase in data size for large $\hat{N}_{obs}$. More specifically, each block of a 256-bit message input is converted into a 6144-bit ciphertext output, which means that for every iteration of the PQC-PKE, the data size increases by a factor of~24. 
Therefore, to mitigate this, we replace PKE with KEM and an encryption scheme. We use Crystals-Kyber as a KEM to establish a 256-bit symmetric key every system cycle. Out of the 256-bit key, 128~bits are used as the PQC key, 120~bits as a base nonce, $v'$, and 8~bits are reserved for the counter. The base nonce concatenated with a unique counter at each intermediate encryption ensures a negligible probability of nonce-key pair collision throughout any cycle. We modify the encryption scheme with Ascon encryption for the following reasons:
\begin{enumerate}
\item Ascon is a lightweight authenticated encryption with an associated data (AEAD)~\cite{dobraunig2014ascon} cipher that provides both confidentiality and built-in authentication with low computational and memory overhead. 
\item Ascon's sponge-based design~\cite{dobraunig2014ascon} supports efficient encryption of arbitrary-length messages without expanding the ciphertext beyond a fixed authentication tag.
\item The Ascon design facilitates the efficient implementation of side-channel countermeasures (e.g., low-degree S-box, levelled protection that narrows the protected region)~\cite{mirigaldi2025quest}.
\end{enumerate}
So the encryption schemes now become $\xi_i' \in \{\text{OTP, AES, Ascon}\}$. Also, since the $\xi'$ does not contain any PKE, the re-encryption attack~\cite {cryptography7040049}, which is associated with a PKE, is now eliminated. The comparison in the execution time of the $\text{HOQS+}$ with respect to the $\text{HOQS}$ is presented in Section~\ref {Sec: results}.

\section{Finite-Key QKD Implementation}
\label{Sec: finiteQKD}

We implement the QKD primitive within our hybrid system, using tight finite-key security frameworks.
The QKD primitive implements an entanglement-based BBM92 protocol with complete finite-key extraction given according to the works of~\cite{tomamichel2017largely,lim2021security, mannalath2025sharp}.  

\subsection{Experimental realization}
In~\cite{tomamichel2017largely}, the authors mention certain assumptions under which they build the finite-key security framework. The same model is used by later developments~\cite{lim2021security, mannalath2025sharp} of the same framework. We report that our experimental implementation of the BBM92 protocol complies with all necessary assumptions.   
Our optical setup employs a spontaneous parametric down-conversion source producing polarization-entangled photon pairs, each photon at 810~nm wavelength. 
True randomness in the basis choice is ensured by 50:50 beam splitters, and complementarity between measurement bases is maintained using half-wave plates.
Detection outcomes include $\{0, 1, \phi\}$, where no-detection events ($\phi$) are recorded as~$-1$ and discarded by processing $0,1$ and $-1$ identically. In this way, the basis information is not revealed, preventing side-channel leakage as described in~\cite{lydersen2010hacking}. Finally, classical post-processing communication is authenticated using an IT secure MAC derived from a subset of the PSK. As a result, the QKD primitive inherently satisfies the finite-key assumptions formulated in~\cite{tomamichel2017largely}, thus reducing the number of explicit assumptions required when adopting the extended composable-security model of~\cite{mannalath2025sharp}. We follow exactly the QKD setup described by~\cite{rani2025combined}, and the reader is referred to that work for complete details of our BBM92 QKD implementation.

\subsection{Finite-key security framework}
In the works of~\cite{lim2021security} and~\cite{mannalath2025sharp}, the authors extend the framework given in~\cite{tomamichel2017largely} by introducing tighter bounds for the parameter estimation error. We follow this framework except for a minor change (incorporating the failure probability in authentication). 
The authentication failure probability (denoted by $\epsilon_{auth}$) of the reconciliation messages (syndrome, hash,
etc.) should
be tightly bound, as any non-negligible chance of successful forgery on the authenticated classical channel directly undermines the composable security of the QKD protocol. A forged authentication tag could allow Eve to insert or modify reconciliation messages, effectively enabling a man-in-the-middle attack. The general composable security parameter, $s$, is defined by $\epsilon_{QKD}~= 10^{-s}$. The following relation should be satisfied:
\begin{align}
\epsilon_{\mathrm{QKD}} \ge
\epsilon_{auth}+\epsilon_{ec}+\epsilon_{pa}+2\epsilon_{pe}
.
\label{eq:security}
\end{align}
Here, $\epsilon_{auth} = q2^{-p}$, $p$ is the length of the binary MAC tag, and $q$ is the number
of authentication tags generated in the QKD reconciliation protocol. Adopting from ~\cite{lim2021security},  $\epsilon_{ec} = 2^{-t}$ denotes the error correction failure probability, where $2^{-t} = 10^{-(s+2)}$. The term $\epsilon_{pa}$ denotes the failure probability in privacy amplification, which is expressed as,
\begin{align}
\label{eq:error_pa}
\epsilon_{pa} = \frac{1}{2}\sqrt{
2^{-(N-n)[1-h_2(\delta+\nu)]+r+t+l}},
\end{align}
where $h_2(x)$ is the binary entropy function, $N$ is the total length of raw bits, of which $n$ bits are used to estimate the quantum bit error rate (QBER). 

Note that the following terminology will be used henceforth. We use the terms $\alpha$, $\beta$, and $\gamma$ to define the computed QBER in the $n$ bits, the true QBER in the $N-n$ bits, and the total QBER in the total $N$ bits. The term $\delta~(\leq 0.11)$ denotes a threshold value tolerated for the computed QBER. If $\alpha$ is above $\delta$ the reconciliation process is aborted   (see Appendix~\ref{qber_estimate_appendix}). The variable  $\nu$ ($0<\nu\leq1/2-\delta$) denotes the maximum estimate of deviation between the true QBER and the threshold, with a confidence of $1-\epsilon_{pe}$. We define $r$ as the length of the syndrome and $l$ as the length of the final extracted QKD key.
The term~$\epsilon_{pe}$ corresponds to the failure probability of the true QBER estimation. We next describe the statistical limit used for $\epsilon_{pe}$.

\subsection{Statistical bounds in QBER estimation failure}
\label{sec: qber failure}
The work of~\cite{lim2021security} extended the proof of~\cite{tomamichel2017largely} by replacing the single Serfling deviation bound with a two-term expression combining Serfling's and a hypergeometric probability bound (Hush and Scovel's inequality). This refinement tightens the estimation of the true QBER failure probability upper bound, denoted by $\epsilon_{pe}^{\text{Serf}}$, as, 
\begin{align}
\label{eq:pe_serf}
    \epsilon_{pe}^{\text{Serf}}= \sqrt{\theta_1+\theta_2}, 
\end{align}
where
$$
\theta_1= \exp \Bigg(-\frac{2Nn\mu^2}{N-n+1}\Bigg), 
$$
$$
\theta_2 = \exp (-2\Gamma_{N(\delta+\mu)}[((N-n)(\nu-\mu))^2-1]),
$$
$$
\Gamma_{N(\delta+\mu)} = \frac{1}{\lfloor N(\delta+\mu)\rfloor+1}+\frac{1}{N-\lfloor N(\delta+\mu)\rfloor+1},
$$
and $0<\mu<\nu$ and $\lfloor.\rfloor$ denotes the floor function.
To further tighten $\nu$, we adopt the recent and tighter bound given by~\cite{mannalath2025sharp}, where the authors introduce two formulations for the total QBER estimation: (i)~an analytic relaxed-Chernoff bound that yields an algebraic closed-form one-sided confidence interval for the total QBER, and (ii)~an exact Clopper--Pearson (CP) construction that numerically inverts the hypergeometric cumulative distribution function to obtain an exact bound for the total QBER. 
The analytic relaxed-Chernoff upper bound on the total QBER with the failure probability, at most, $\epsilon_{pe}^{\mathrm{Chern}}$ is related to the threshold as
\begin{align}
\label{eq:gamma+}
\Gamma^+_{n,\epsilon_{pe}^{\text{Chern}}}(\delta)
=\frac{3\kappa+(1-2\kappa)\delta
+3\sqrt{\kappa(\kappa+\delta-\delta^2)}}{1+4\kappa},
\end{align}
 
$$\text{where, }\kappa = \frac{2}{9n}\ln\frac{1}{\epsilon_{pe}^{\text{Chern}}}.$$
Hence, the relation between the deviation, $\nu$, the upper bound on the total QBER and the threshold is given by (for derivation see Appendix~\ref{gamma+nu_appendix})
\begin{align}
\label{eq:gamma_chern}
\nu=\frac{N(\Gamma^+_{n,\epsilon_{pe}^{\text{Chern}}}(\delta)-\delta)}{N-n}.
\end{align}

 Alternatively, with the CP construction, we can obtain a minimum $\nu (\geq0)$ such that the failure probability of the event $\alpha\leq\delta$ when the minimum estimate of total error $K = N(\delta+\nu)-n\nu\geq N\delta$, is at most $\epsilon_{pe}^{\text{CP}}$.
Since $\lfloor n\alpha\rfloor$ follows a hypergeometric distribution, the relation of $\epsilon_{pe}^{\text{CP}}$ with smallest $\nu$ is given by

\begin{align}
\label{eq: eps_CP}
\epsilon_{pe}^{\text{CP}}=\sum_{x=0}^{\lfloor\delta n\rfloor} \frac{\binom{K}{x} \binom{N-K}{n-x}}{\binom{N}{n}}.
\end{align}
The detailed derivation of {Eq.~\ref{eq: eps_CP}} is given in Appendix~\ref{K_nu_appendix}.

\subsection{Optimization}
The security framework of~\cite{lim2021security} and~\cite{mannalath2025sharp}, performed optimization of the vector, $\vec{x} = (l/N,n/N,\nu,\mu)$ and $\vec{x} = (l/N,n/N)$ for a given known $\delta$, respectively. However, in a real QKD implementation, it is not optimal to choose a threshold prior to computing $\alpha$, rather, to set $\delta$ equal to $\alpha$ (see Appendix~\ref{qber_estimate_appendix}).
In our system, we fix the parameters $N$, $n = N/2$, $p$, $q$ and $r$.
Our MAC tag keys come from the PSK.
In~\cite{lim2021security}, the authors set
$r = r'$, where $r' = 1.19(N-n) \times h_2(\delta)$; however, we choose a larger fixed $r\geq r'$, because, although using $r'$ can increase the QKD key rate, it also increases the probability that the LDPC-based error correction fails. This failure corresponds to a detectable failure, where the decoder cannot find a valid codeword or a parity check fails, distinct from the undetected failure captured by $\epsilon_{ec}$. Therefore, there exists a trade-off between the number of times the QKD protocol aborts because of this failure and the achievable finite key rate per session (a QKD session involves one complete post-processing process). We choose a larger fixed $r$ to reduce the number of times the hybrid system aborts. The fixed parameters $N$ and $n$ are chosen as reasonable heuristic values that are large enough to permit final QKD key extraction, yet small enough to avoid impractically long acquisition times. With these fixed parameters, we optimize the length $l$ (finding the value of $\nu$, that leads to the maximum $l$). The values of $N$ and $l$ are then inputs to the privacy amplification procedures. The optimization problem for the QKD key length becomes,
\begin{align}
\label{eq:opt_prob}
    \max_{\vec{x}=(\nu)} ~&\{l |~2\epsilon_{pe}+\epsilon_{ec}+\epsilon_{pa}+\epsilon_{auth}\le \epsilon_{QKD}\}, \\
l>0, &~\nu \in (0,1/2-\delta] \nonumber.
\end{align}
We perform this optimization for a fixed $s$ and the $\epsilon_{pe}$ is computed with all three QBER estimation failure probabilities ($\epsilon_{pe} \in \{\epsilon_{pe}^{\text{Serf}},\epsilon_{pe}^{\text{Chern}},\epsilon_{pe}^{\text{CP}}\}$) (see Appendix~\ref{algorithms_appendix}). We computed $\epsilon_{pe}^{\text{Serf}}$ directly from Eq.~\ref{eq:pe_serf} with the optimization that also runs overs $\mu$ along with $\nu$. While for $\epsilon_{pe}^{\text{Chern}}$ and $\epsilon_{pe}^{\text{CP}}$, we perform a binary search over different values of $\epsilon_{pe}^{\text{Chern}}$ and $\epsilon_{pe}^{\text{CP}}$. We present, in Fig.~\ref{fig:eff_key_qber}$\text{(b)}$, the finite QKD key rate results obtained from the above optimization problem by employing different $\epsilon_{pe}$ given in this section. We also show the optimization times, the estimated deviation, $\nu$ and the extracted positive key rate with different $\epsilon_{pe}$ bounds for an example $\hat{N}_{obs}$, $\alpha$ and $s$ in Table~\ref{tab: security_opt}.

\section{Results\label{Sec: results}}
\begin{figure}[!ht]
\centering

    \begin{subfigure}[b]{\linewidth}
     \begin{overpic}[width=0.9\linewidth]{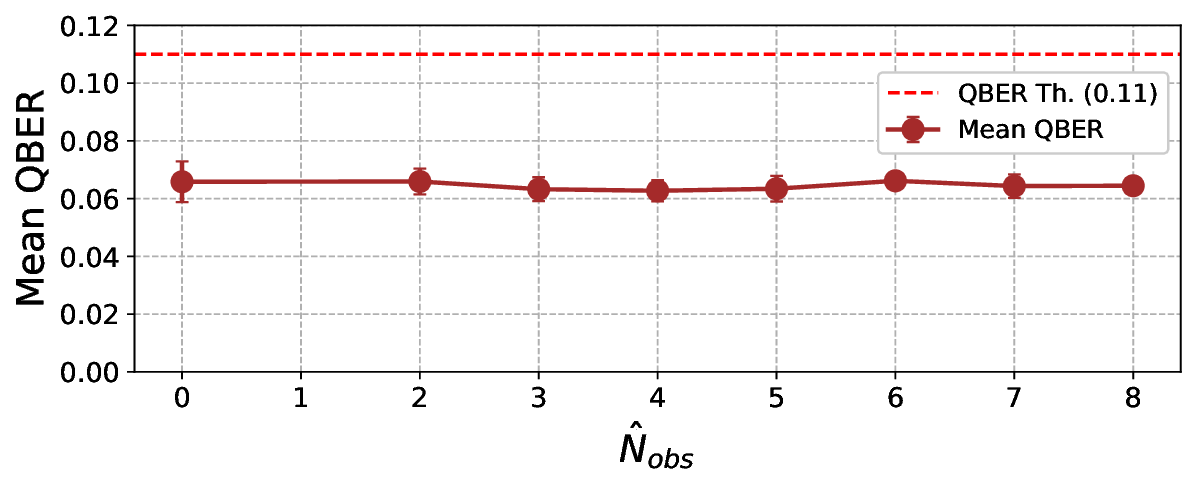}
         \put(-2,38){(a)} 
    \end{overpic}
        \label{fig:qber}\\
        
    \end{subfigure}\\
    
    \begin{subfigure}[b]{\linewidth}
    \begin{overpic}[width=0.9\linewidth]{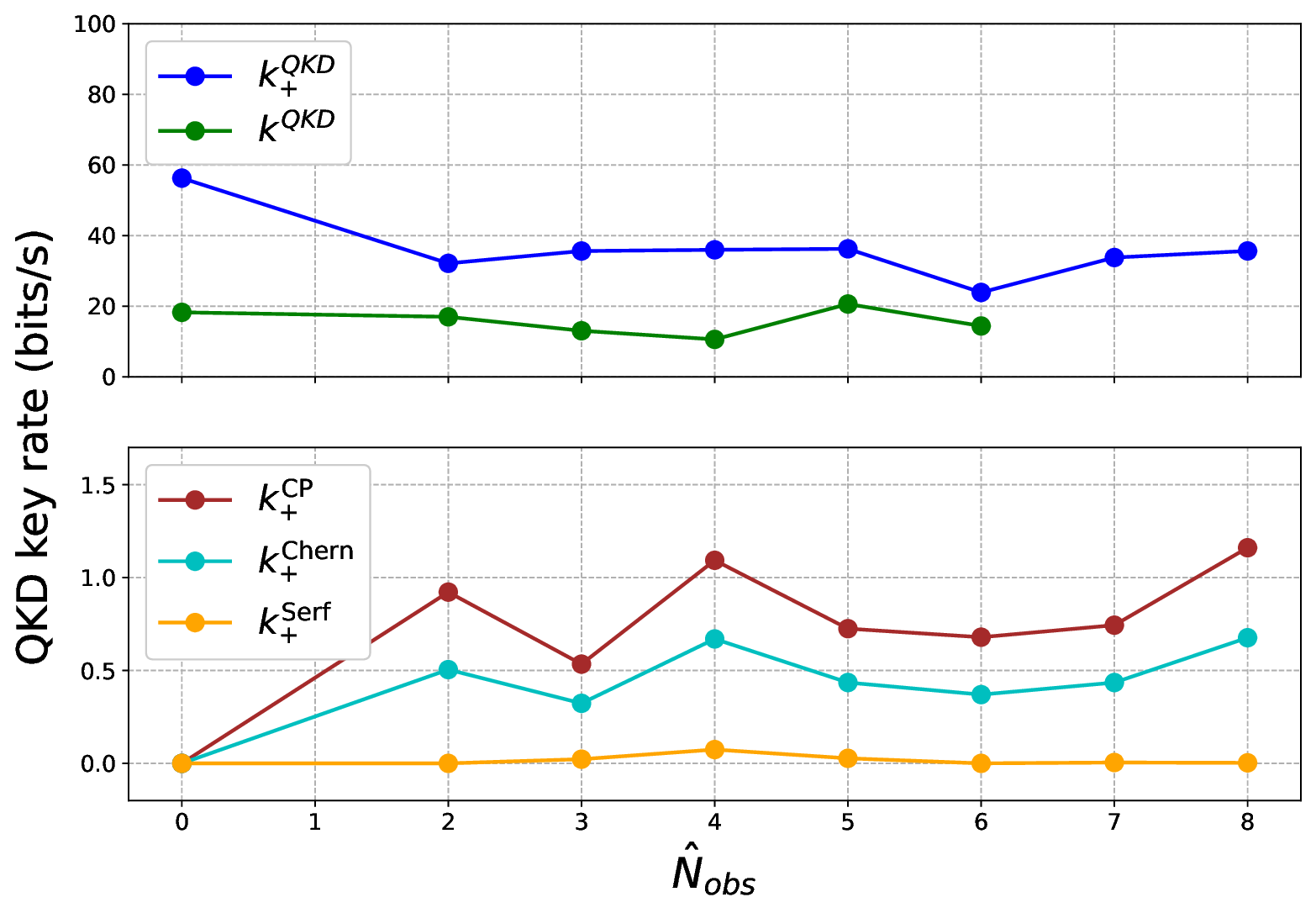}
         \put(-2,65){(b)} 
    \end{overpic}
        \label{fig:key_rates}
    \end{subfigure}
    \caption{QBER (in (a)) and the corresponding QKD key rates (bits/s) (in (b)) for different $\hat{N}_{obs}$. Here, $k^{\mathrm{QKD}}$ and $k_{+}^{\mathrm{QKD}}$ are asymptotic QKD key rates for both the HOQS and the HOQS+, respectively, while $k^{\text{Chern}}_{+}$, $k^{\text{CP}}_{+}$ and $k^{\text{Serf}}_{+}$, represent finite QKD key rates of the $\text{HOQS+}$ per system cycle, corresponding to optimizations using $\epsilon_{pe}^{\text{Chern}}$, $\epsilon_{pe}^{\text{CP}}$ and $\epsilon_{pe}^{\text{Serf}}$, respectively. `QBER Th.' marks the QBER threshold (0.11), and `Mean QBER' is the average of the computed QBER over 10 system cycles. $k^{\mathrm{QKD}}$ was $0$ for $\hat{N}_{obs}>6$ as the HOQS could not successfully complete a single run beyond $\hat{N}_{obs}=6$. Here, the security parameter, $s = 6$.}
    \label{fig:eff_key_qber}
\end{figure}

We implemented both systems in a laboratory setting where the QKD keys were established between Alice and Bob, who were separated by a~1.5~m free-space channel. The BBM92 protocol with entangled photons (entangled in the polarization degree of freedom) established the QKD keys in real-time. The raw time tags were recorded by a Quantum Machine (a post-processing device with an in-built FPGA). The time synchronization scheme detailed in~\cite{rani2025obfuscatedquantumpostquantumcryptography} was then used to synchronize the time tags corresponding to the arrival time of the entangled photons at Alice's and Bob's detectors. The QKD post-processing steps were then performed on the time tags using identical laptops with~16~GB RAM and~2.9~GHz clock speed GPU for both parties to generate fresh QKD keys.~$\lceil\hat{N}_{obs}/2\rceil$ QKD sessions were executed within a single hybrid system cycle. Subsequently, these QKD keys were used to encrypt an intermediate ciphertext~$(c_i)$ with the OTP encryption scheme. Within the hybrid system, HOQS+, fresh PQC keys were established between Alice and Bob in each cycle via the execution of Kyber KEM. The PQC keys were used with Ascon encryption. Additionally,~$256$~bits of PSK were used with the AES encryption scheme, while 61~bits were used for channel authentication and $\hat{N}_{obs}$~bits for IS encryption. We used a $128$-bit base nonce, $v$, a $120$-bit sID, and an $8$-bit counter. In order to demonstrate the improvements in the scalability of the proposed $\text{HOQS+}$ compared to $\text{HOQS}$, we executed both systems with a fixed sample text message of size~102~bytes. The scalability improvement is demonstrated both in terms of the total processing time and the asymptotic key rates of the extracted QKD key per cycle of both systems. We executed~10 cycles of both systems with each $\hat{N}_{obs}$ and observed an average QBER of $0.0644\pm0.0037$  in the QKD primitive - see Fig.~\ref{fig:eff_key_qber}$\text{(a)}$.   

\begin{figure*}[!ht]
    \centering
    \includegraphics[width=0.82\linewidth]{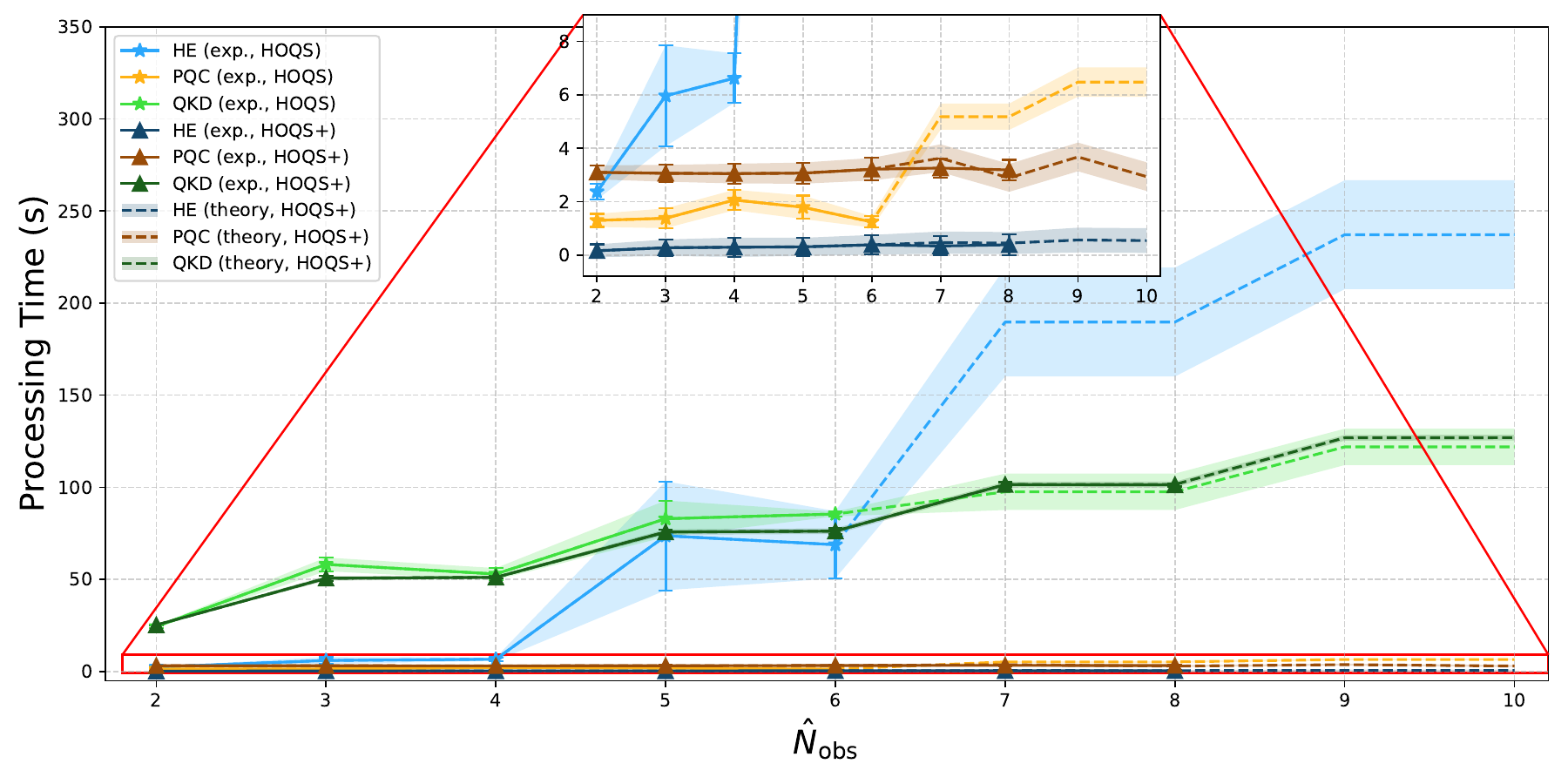}
    \caption{Processing times (s) of different primitives in both the HOQS and the HOQS+ for~one cycle. We plot the observed experimental mean processing times with solid lines, while the projected processing time for larger $\hat{N}_{obs}$ is shown with dashed lines. The `HE' in the label corresponds to HE, the `exp' corresponds to the experimentally observed times, while theory corresponds to the projected fit of the observed data. The light shade plots represent the $\text{HOQS}$, while the corresponding darker shade for the $\text{HOQS+}$. The significant difference in the scaling of the two systems comes from the comparison of the processing times of the HE for the two systems. While the processing time of the HE in the $\text{HOQS}$ scales much faster than the factor of $\hat{N}_{obs}$, such that the system experienced timeout exceptions at large $\hat{N}_{obs}$, the processing time of the HE and its scaling in the $\text{HOQS+}$ are almost constant in $\hat{N}_{obs}$.}
    \label{fig:total_exec_time}
\end{figure*}
A hybrid system cycle began with the sharing of the encrypted instruction sequence, followed by QKD post-processing, the establishment of the PQC key, and then the HE of the sample message. The total processing time was measured as the sum of the processing times of the QKD, PQC, and HE primitives combined over 10 cycles. The QKD primitive, in turn, involves the time taken in the synchronization of time tags, basis sifting, error correction, authentication, and privacy amplification processes. The PQC primitive performs symmetric key sharing via KEM, and the HE primitive executes cascaded encryption of a message to be communicated.

In Fig.~\ref{fig:eff_key_qber}~$\text{(b)}$, the asymptotic QKD key rates for both systems are represented as $k^{\mathrm{QKD}}$ and $k_\mathrm{+}^{\mathrm{QKD}}$, respectively, and the finite QKD key rates with $\text{HOQS+}$ are represented as $k^{\text{Chern}}_{\mathrm{+}}$, $k^{\text{CP}}_{\mathrm{+}}$ and $k^{\text{Serf}}_{\mathrm{+}}$. We derive these key rates by adding the total extracted QKD keys in 10 continuous runs of the hybrid system cycles, divided by the total processing time taken by the HOQS and HOQS+, respectively. Comparing $k^{\mathrm{QKD}}$ and $k_{\mathrm{+}}^\mathrm{{QKD}}$, in Fig.~\ref{fig:eff_key_qber}$\text{(b)}$, shows a consistent improvement in the extracted QKD key across all values of $\hat{N}_{obs}$. This is due to the reduction in the processing times of the $\text{HOQS+}$ relative to the $\text{HOQS}$ (shown in Fig.~\ref{fig:total_exec_time}), more specifically, due to the modification of the HE primitive. The $\text{HOQS}$, in fact, could not extract any QKD keys for $\hat{N}_{obs}>6$ due to timeout exceptions (a timeout exception is an error condition raised when the elapsed time for a blocking or asynchronous operation exceeds its configured timeout threshold). 

We performed the optimization mentioned in Eq.~\ref{eq:opt_prob} for different $\epsilon_{pe} \in \{\epsilon_{pe}^{\text{Serf}},\epsilon_{pe}^{\text{Chern}},\epsilon_{pe}^{\text{CP}}\}$ with fixed parameters, $s=6$, $N=2\times10^4, p=61, r= 5000$, and $q=1$. We chose $s=6$ for comparison of $\epsilon_{pe}^{\text{Chern}}$ and $\epsilon_{pe}^{\text{CP}}$ with $\epsilon_{pe}^{\text{Serf}}$, as, the $\epsilon_{pe}^{\text{Serf}}$ could not extract any positive key rate at $s>6$ with our choice of $N$ and observed QBERs. From Fig.~\ref{fig:eff_key_qber}$\text{(b)}$, we see that the bound given in~\cite{mannalath2025sharp} extracts a much larger QKD key rate than the bound given in~\cite{lim2021security} and improves security (Table ~\ref{tab: security_opt}). We see that, in alignment with the results in~\cite{mannalath2025sharp}, the exact Clopper-Pearson solution outperforms the analytical solution using the relaxed additive Chernoff bound in terms of higher key rates, however takes a much longer time to optimize. 
Using the bound, $\epsilon_{pe}^{\text{Serf}}$ gives an overestimated value of $\nu$ and hence a much lower key rate. Also we observed that setting $s=9$ at $\alpha = 0.0627$ and $N=2\times10^4$, the system could not extract any key using $\epsilon_{pe}^{\text{Serf}}$, while, for $\epsilon_{pe}^{\text{Chern}}$ and $\epsilon_{pe}^{\text{CP}}$ the key rate ($l/N$) were $0.015$ and $0.062$, respectively. 
\begin{table}[]
    \centering
    \caption{Eq.~\ref{eq:opt_prob} is optimized at different $\epsilon_{pe}\in \{\epsilon_{pe}^{\text{Serf}},\epsilon_{pe}^{\text{Chern}},\epsilon_{pe}^{\text{CP}}\}$ for an example execution where, $s=6$, $N=2\times10^4$, $\hat{N}_{obs}=4$, and $\alpha=0.0627$. The $\epsilon_{pe}^{\text{Serf}}$ loosely bounded the hypergeometric tail, therefore overestimated the $\nu$ in all the cases. We see a trade-off in execution time and $l/N$ between $\epsilon_{pe}^{\text{Chern}}$ and $\epsilon_{pe}^{\text{CP}}$.}
    \vspace{0.5 em}
    \begin{tabular}{|c|c|c|c|}
    \hline
    \textbf{$\epsilon_{pe}$} &$\nu$& $l/N $ & time (s)\\
        \hline
         $\epsilon_{pe}^{\text{Serf}}$ & 0.043
 & 0.003 & 12.51\\
         $\epsilon_{pe}^{\text{Chern}}$ & 0.023
 & 0.027 & 397.63\\
         $\epsilon_{pe}^{\text{CP}}$ & 0.006 & 0.066 & 415.56\\
         \hline
    \end{tabular}
    
    \label{tab: security_opt}
\end{table}

In Fig.~\ref{fig:total_exec_time}, we show the individual processing times of each primitive (the QKD, the PQC, and the HE) of the hybrid systems. We plot the observed mean and standard deviation of the processing times as solid lines, and their projected values for higher $\hat{N}_{obs}$ as dotted lines. The projection is obtained using a hybrid regression model that fits the observed data over a baseline scaling function proportional to $\lceil\hat{N}_{obs}/2\rceil$. As can be seen from the figure, the processing time of the HE in the $\text{HOQS}$ scales much larger than the factor $\lceil\hat{N}_{obs}/2\rceil$. This is because the size of $c_i$ also increased rapidly after each consecutive encryption in the $\text{HOQS}$. This impacted the scaling the $\text{HOQS}$ for higher $\hat{N}_{obs}$ values and caused timeout exceptions beyond $\hat{N}_{obs}=6$. In contrast,  with the proposed optimizations in HE given in Sec.~\ref{Sec: system_model}, in the $\text{HOQS+}$, the processing time of the HE along with the projection is now almost constant in $\hat{N}_{obs}$. The processing times and projections (comparing the dark and light shades of green) in the QKD primitive (which is linear in $\lceil\hat{N}_{obs}/2\rceil$) remain almost the same in both systems. While for the PQC primitive, we see a slight constant increase in the processing times between HOQS and HOQS+ because of the additional encapsulation and decapsulation step in the latter.

\section{Conclusion\label{Sec: conclusion}}
In this work, we presented an improved design for a hybrid QKD-PQC system, scaling with the parameter $\hat{N}_{obs}$. As this parameter increases, more unique instruction sequences can be formed, informing the receiver how to decrypt incoming messages. A large $\hat{N}_{obs}$ implies higher resource requirements for an adversary to compromise the hybrid system. In the case of a simultaneous side-channel attack on the QKD and PQC primitives, this system is argued to retain the confidentiality of a message in any pragmatic real-world setting. 

In addition, we implemented a realistic and robust QKD primitive with enhanced security in a finite-key-rate security framework. We believe that this is the first implementation of a new finite-key security framework recently announced that offers dramatic improvement in the QKD key rates.  We also showed that $\text{HOQS+}$ achieved higher asymptotic QKD key rates compared to $\text{HOQS}$, and $\text{HOQS+}$ also maintains positive finite QKD key rates even when a very large number of instruction sequences are allowed (Fig.~\ref{fig:eff_key_qber}$\text{(b)}$).   We modified the HE schemes to reduce the execution overhead and experimentally verified the processing time improvements of the $\text{HOQS+}$ over the $\text{HOQS}$ (Fig.~\ref{fig:total_exec_time}). 

In our design, the encryption mechanism and the key generation scheme within the PQC component were deliberately decoupled to further mitigate potential information leakage. Consequently, Ascon encryption was used alongside OTP and AES. Within an HE setting, where multiple cascaded encryptions are applied to the same message, the lightweight nature of Ascon minimizes performance overhead. Furthermore, authentication ensures the integrity of the ciphertext; therefore, Ascon's built-in authentication capability complements AES in the counter mode and OTP, which inherently lack authentication. Generating a nonce once per cycle and incorporating session IDs and the key-counter in both Ascon and AES ensured the uniqueness of the counter block (with very low collision probability). Additionally, concatenating the nonce at the end of the ciphertext, as opposed to the standard practice, meant that the intermediate ciphertext size remained constant over subsequent encryptions.

The deployment of hybrid QKD-PQC schemes is particularly valuable in environments where no single primitive can be guaranteed to be completely secure against side-channel attacks. Such systems are especially relevant to defence applications, where long-term communication confidentiality is critical, and it is not clear which primitive should be deployed alone. The use of PSK, for directly encrypting the message, is not a long-term solution, and lacks any key growth algorithm; therefore, optimized use of a preshared key to encrypt a small instruction that defines the configuration of a hybrid system is a better option and can dramatically improve confidentiality of the message even in the case of simultaneous side-channel attacks in all key generation primitives. We believe that all QKD systems deployed in the future will involve some form of hybrid encryption similar to that outlined here.

Future work could focus on the integration of data-based encryption, where the encryption strategy is determined by the sensitivity of the data that forms a subset of the message. This may prove to be a better use of the generated keys with different key generation primitives. Additionally, integrating hybrid cryptographic schemes with a secret sharing scheme~\cite{beimel2011secret} and physically unclonable functions (PUFs)~\cite{gao2020physical} may prove fruitful. The combination of hybrid QKD-PQC with other physical attributes, such as the location of transceivers, is likely to lead to further enhanced security outcomes.

\section*{Acknowledgment}
The authors thank Dr. Dushy Tissainayagam from Northrop
Grumman Australia for useful discussions. This research
has been carried out as a project co-funded by Northrop
Grumman Australia, and the Defence Trailblazer Program,
a collaborative partnership between the University of Adelaide
and the University of New South Wales, co-funded by the
Australian Government, Department of Education.

{\small
\bibliographystyle{IEEEtran}
\bibliography{IEEEabrv,biblio}
}

\newpage
\appendix
\subsection{QBER estimation}
\label{qber_estimate_appendix}
Let Alice and Bob share the $N$ bits before post-processing in the QKD as $(X, V)$ and $(Y, W)$, of sizes 
$|V|=|W|=n$, and $|X|=|Y|=(N-n)$. Hence, $\alpha = |V\otimes W|/k$, $\beta = |X\otimes Y|/n$ and $\gamma~(= K/N)$, and therefore 
\begin{equation}
\label{eq. total error rate}  
    N\gamma = (N-n)\beta+n\alpha. \tag{a}
\end{equation}
The goal is to upper bound the unknown value of $\beta$, using the computed value of $\alpha$ with a confidence of $(1-\epsilon_{pe})$. The probability of failure in the estimation of $\beta$ is defined by the probability of the event,
\begin{equation}
\label{eq: failure pe}
    \text{Pr}[\underbrace{\alpha\leq\delta \text{ and } \beta\geq \delta+\nu}_{\text{a bad event}}], \tag{b}
\end{equation}
where, if $\alpha>\delta$, then the QKD reconciliation protocol is aborted.
Consistent with the existing literature~\cite{mannalath2025sharp, tomamichel2017largely} and~\cite{lim2021security}, we can set a $\delta$  between 0 and 0.11. However, we set $\alpha = \delta$, as an optimal strategy because, if a value of $\delta$ is less than $\alpha$ then it leads to more abort conditions, while if $\delta$ is more than $\alpha$ then it leads to too pessimistic estimation of $\beta$ and hence too low QKD key rates.

\subsection{The relation between $\Gamma^+$ and $\nu$}
\label{gamma+nu_appendix}
Here, we derive Eq.~\ref{eq:gamma_chern} by mapping it to Eq.~6 of~\cite{mannalath2025sharp}. In the work~\cite{mannalath2025sharp}, Eq.~6 is given as
\[
    q^{th} = \frac{N\Gamma^+-n\delta}{N-n},
    \]
where, $q^{th}$ is the maximum estimate of the true QBER,
\begin{equation}
\tag{c}
\begin{aligned}
    \implies &\delta+\nu = \frac{N\Gamma^+-n\delta}{N-n},\\
    \implies&\nu = \frac{N(\Gamma^+-\delta)}{N-n}.
    \end{aligned}
\end{equation}

\subsection{The relation between $K$ and $\nu$}
\label{K_nu_appendix}
We now justify the reason for using $\mathcal{CP}^{+}$ to estimate $\nu$, mentioned in Eq.~\ref{eq: eps_CP}, in the main text. In an exact CP construction that numerically inverts the hypergeometric cumulative distribution function to obtain an exact minimum value of $\gamma$. In~\cite{mannalath2025sharp}, the term $\mathcal{CP}^+$ describes the tight estimation of $\gamma$. However, we use it to tightly estimate the deviation $\nu$ via the following transformation,
\begin{equation}
\label{eq: CP+}
\mathcal{CP}^+_{n,\epsilon_{pe}^{\text{CP}}}(\delta)
= \min_{\gamma\geq \delta}\Big\{\gamma \mid \Pr[\alpha\le \delta\mid N,K,n]\le\epsilon_{pe}^{\text{CP}}\Big\}. \tag{d}
\end{equation}
From Eq.~\ref{eq. total error rate} and Eq.~\ref{eq: CP+}, we minimize
\begin{equation}
\label{eq: K range}
\tag{e}
\begin{aligned}
\gamma=\frac{n-N}{N}\beta+\frac{n}{N}\alpha&\geq \delta, \nonumber\\
\text{substituting $\alpha$ and $\beta$ from Eq.~\ref{eq: failure pe},} \nonumber\\
\implies N\gamma=(N-n)(\delta+\nu)+n\delta&\geq N\delta, \nonumber\\
\implies K=N(\delta+\nu)-n\nu&\geq N\delta,  \nonumber\\
\end{aligned}
\end{equation}
hence, Eq.~\ref{eq: CP+} becomes 
\begin{equation}
\label{eq: CP++}
\mathcal{CP}^+_{n,\epsilon_{pe}^{\text{CP}}}(\nu)
= \min_{\nu\geq 0}\Big\{\nu \mid \underbrace{\Pr[\alpha\le \delta\mid N,K,n]}_{\text{bad event}}\le\epsilon_{pe}^{\text{CP}}\Big\}, \tag{f}
\end{equation}
and the maximum probability of the bad event can be given as 
\begin{equation}
\label{eq: bad_bound}
\Pr\underbrace{[\alpha\le \delta\mid N,K,n]}_{\text{bad event}}=\epsilon_{pe}^{\text{CP}}, \tag{g}
\end{equation}
when, $ K = N(\delta+\nu)-n\nu.$

\subsection{Algorithms for optimization}
\label{algorithms_appendix}
The pseudo-codes, Algorithm~\ref{algo: eps_pa},~\ref{algo: eps_pe},~\ref{algo: eps_pe_chern},~\ref{algo: eps_cp} and~\ref{algo: optimize}   explain the operations associated with the privacy amplification (Eq.~\ref{eq:error_pa}), parameter estimation with improved Serfling's bound (Eq.~\ref{eq:pe_serf}), analytical solution to Chernoff's bound (Eq.~\ref{eq:gamma+}), exact solution with CP construction (Eq.~\ref{eq: eps_CP}), and the optimization of $l$ (Eq.~\ref{eq:opt_prob}), respectively. For example, when optimization with Serfling's bound, one runs Algorithm~\ref{algo: optimize} where Algorithm~\ref{algo: eps_pe} and Algorithm~\ref{algo: eps_pa} are called within.

\begin{algorithm}[]
\DontPrintSemicolon
\caption{Privacy amplification failure probability $\epsilon_{pa}$}
\label{algo: eps_pa}
\KwIn{$l$, $t$, $\nu$, $\delta$, $r$, $n$}
\KwOut{$\epsilon_{pa}$}

$H \gets h_2(\delta + \nu)$ \tcp*[r]{Binary entropy}

$E \gets -n (1 - H) + r + t + l$\;

$\epsilon_{pa} \gets \dfrac{1}{2} \sqrt{2^{\,E}}$\;

\Return $\epsilon_{pa}$\;

\end{algorithm}

\begin{algorithm}[]
\DontPrintSemicolon
\caption{Serfling–hypergeometric QBER estimation failure probability $\epsilon_{pe}^{\text{Serf}}$}
\label{algo: eps_pe}
\KwIn{$\nu-\mu$, $\delta$, $\mu$, $N$, $n$}
\KwOut{$\epsilon_{pe}$}
    $m_{\text{err}} \gets m(\delta + \mu)$\;
    $a \gets \left\lfloor m_{\text{err}} \right\rfloor + 1$\;
    $b \gets m - \left\lfloor m_{\text{err}} \right\rfloor + 1$\;
   $\gamma \gets \dfrac{1}{a} + \dfrac{1}{b}$\;

$\theta_1 \gets \exp\!\left( -\dfrac{2 N n \mu^2}{\,N-n + 1\,} \right)$\;

$\theta_2 \gets \exp\!\left( -2 \gamma \left( ((N-n)(\nu-\mu))^2 - 1 \right) \right)$\;

$\epsilon_{pe} \gets \sqrt{E_1 + E_2}$\; \tcp*[r]{Using Eq.~\ref{eq:pe_serf}}

\Return $\epsilon_{pe}$\;
\end{algorithm}
\begin{algorithm}[]
\DontPrintSemicolon
\caption{Chernoff-Based Parameter Estimation Failure Probability $\epsilon_{pe}^{\text{Chern}}$}
\label{algo: eps_pe_chern}
\KwIn{$\delta$, $\nu$, $n$, $N$, $\text{tolerance: tol}$}
\KwOut{$\epsilon_{pe}$}

$\text{target} \gets \nu$\;

$y_{\text{low}} \gets 0$; $y_{\text{high}} \gets 1000$\; 

\For{$i \gets 1$ \KwTo $2000$}{
    $y_{\text{mid}} \gets (y_{\text{low}} + y_{\text{high}})/2$\;
    $\kappa \gets 2y_{\text{mid}}/(9n)$\;
    $\gamma \gets \frac{3\kappa + (1-2\kappa)\delta +3\sqrt{\kappa(\kappa+\delta-\delta^2)}}{1 +4\kappa}$\; \tcp*[r]{Using Eq.\ref{eq:gamma+}}
    $\nu_{\text{pred}} \gets \dfrac{N(\gamma - \delta)}{N - n}$\;
    \If{$\left|\nu_{\text{pred}} - \text{target}\right| < \text{tol}$}{
        \textbf{break}\;
    }

    \uIf{$\nu_{\text{pred}} < \text{target}$}{
        $y_{\text{low}} \gets y_{\text{mid}}$ \tcp*[r]{Need smaller $\epsilon$}
    }
    \Else{
        $y_{\text{high}} \gets y_{\text{mid}}$ \tcp*[r]{Need larger $\epsilon$}
    }
}

$\epsilon_{pe} \gets e^{-y_{\text{mid}}}$\;

\Return $\epsilon_{pe}$\;

\end{algorithm}

\begin{algorithm}

\DontPrintSemicolon
\caption{Exact CP-based Parameter Estimation Failure Probability $\epsilon_{pe}^{\text{CP}}$}
\label{algo: eps_cp}
\KwIn{$\delta$, $\nu$, $N$, $n$}
\KwOut{$\epsilon_{pe}$}

$K\gets \mathrm{round}(N(\delta + \nu)-n\nu)$ \tcp*[r]{Using Eq.~\ref{eq: K range}}

$K \gets \min(\max(K,0),\, N)$ \tcp*[r]{Ensure $0 \le K \le N$}

$x_{\mathrm{obs}} \gets \mathrm{round}(\delta n)$\;

$\epsilon_{pe} \gets \mathrm{HypergeomCDF}(x_{\mathrm{obs}}; N, K, n)$\; \tcp*[r]{Using Eq.~\ref{eq: bad_bound}}

\Return $\epsilon_{pe}$\;
\end{algorithm}

    \begin{algorithm}[]
\DontPrintSemicolon
\caption{Optimize Secret Key Length $l$ under Finite-Key Constraints}
\label{algo: optimize}
\KwIn{$s$, $\delta$, $N$, $p$, $q$, \texttt{pe\_type}}
\KwOut{$l_{\max}$, $\nu^\star$, $\mu^\star$}

$n \gets \lfloor N/2\rfloor$; 
$r \gets 5000$ \tcp*[r]{Syndrome length }

$l_{\max} \gets$ \texttt{None};
$\nu^\star \gets$ \texttt{None};
$\mu^\star \gets$ \texttt{None}\;
$\nu_{\text{range}} \gets \mathrm{Linspace}\bigl(10^{-6},\, 0.5 - \delta - 10^{-6},\, 100000\bigr)$\;
$t \gets \left\lceil (s + 2)\log_2 10 \right\rceil$\;
$\epsilon_{\mathrm{QKD}} \gets 10^{-s}$\;

\ForEach{$\nu \in \nu_{\text{range}}$}{
    \uIf{\texttt{pe\_type} = `serfling'}{
        $\mu_{\text{range}} \gets \mathrm{Linspace}\bigl(10^{-7},\, \nu - 10^{-7},\, 1000\bigr)$\;
        \ForEach{$\mu \in \mu_{\text{range}}$}{
            $\epsilon_{pe} \gets \epsilon_{pe}^{\text{Serf}}(\nu - \mu, \delta, \mu, N, n)$\;
            $\epsilon_{auth} \gets q2^{-p}$\;
            $\epsilon_{ec} \gets 2^{-t}$\;

            $B \gets \epsilon_{\mathrm{QKD}} - \epsilon_{ec} - 2\epsilon_{pe} - \epsilon_{auth}$\;
            \If{$B \le 0$}{
                \textbf{continue}\;
            }

            $H \gets h_2(\delta + \nu)$\;
            $l_{\text{est}} \gets \log_2(4B^2) + n(1 - H) - r - t$\;
            $l \gets \lfloor l_{\text{est}} \rfloor$\;
            \If{$l \le 0$}{
                \textbf{continue}\;
            }

            $\epsilon_{pa} \gets \epsilon_{pa}(l, t, \nu, \delta, r, n)$\;
            $\epsilon_{\text{total}} \gets 2\epsilon_{pe} + \epsilon_{ec} + \epsilon_{pa} + \epsilon_{auth}$\;

            \If{$\epsilon_{\text{total}} > \epsilon_{\mathrm{QKD}} \cdot (1 + 10^{-10})$}{
                \textbf{continue}\;
            }

            \If{$(l_{\max} = \texttt{None})$ \textbf{or} $(l > l_{\max})$}{
                $l_{\max} \gets l$\;
                $\nu^\star \gets \nu$\;
                $\mu^\star \gets \mu$\;
                
            }
        }
    }
    \Else{
        \uIf{\texttt{pe\_type} = `chernoff'}{
            $\epsilon_{pe} \gets \epsilon_{pe}^{\text{chern}}(\delta, \nu, N, n)$\;
        }
        \uElseIf{\texttt{pe\_type} = `cp\_exact'}{
            $\epsilon_{pe} \gets \epsilon_{pe}^{CP}(\delta, \nu, N, n)$\;
        }

        $\epsilon_{auth} \gets q2^{-p}$\;
        $\epsilon_{ec} \gets 2^{-t}$\;

        $B \gets \epsilon_{\mathrm{QKD}} - \epsilon_{ec} - 2\epsilon_{pe} - \epsilon_{auth}$\;
        \If{$B \le 0$}{
            \textbf{continue}\;
        }

        $H \gets h_2(\delta + \nu)$\;
        $l_{\text{est}} \gets \log_2(4B^2) + n(1 - H) - r - t$\;
        $l \gets \lfloor l_{\text{est}} \rfloor$\;
        \If{$l \le 0$}{
            \textbf{continue}\;
        }

        $\epsilon_{pa} \gets \epsilon_{pa}(l, t, \nu, \delta, r, n)$\;
        $\epsilon_{\text{total}} \gets 2\epsilon_{pe} + \epsilon_{ec} + \epsilon_{pa} + \epsilon_{auth}$\;

        \If{$\epsilon_{\text{total}} > \epsilon_{\mathrm{QKD}} \cdot (1 + 10^{-100})$}{
            \textbf{continue}\;
        }

        \If{$(l_{\max} = \texttt{None})$ \textbf{or} $(l > l_{\max})$}{
            $l_{\max} \gets l$\;
            $\nu^\star \gets \nu$\;
    
        }
    }
}

\Return $(l_{\max}, \nu^\star, \mu^\star)$\;

\end{algorithm}

\end{document}